\documentclass[aps,pra,twocolumn,showpacs,groupedaddress]{revtex4} 
\usepackage{graphicx}  
\usepackage{dcolumn}   
\usepackage{bm}       
\usepackage{amssymb}  
\usepackage{amsmath}
\usepackage{float}
\usepackage{enumerate}

\begin{document}
\title{Systems that become $\cal{PT}$ symmetric through interaction}
\author{S. Karthiga$^{1}$, V.K. Chandrasekar$^{2}$, M. Senthilvelan$^{1}$, M. Lakshmanan$^{1}$}
\address{$^1$Centre for Nonlinear Dynamics, School of Physics, Bharathidasan University, Tiruchirappalli - 620 024, Tamil Nadu, India.\\
$^2$Centre for Nonlinear Science \& Engineering, School of Electrical \& Electronics Engineering, SASTRA University, Thanjavur -613 401, Tamilnadu, India.}
 \begin{abstract} 
\par In this article, we consider an interesting class of optical and other systems in which the interaction or coupling makes the systems to be $\cal{PT}$-symmetric. We aim to compare their dynamical behaviors with that of the usual $\cal{PT}$ symmetric systems with intrinsic loss-gain terms. In particular, we focus on the interesting non-reciprocal nature of the $\cal{PT}$ symmetric systems which has a promising application in optical diodes and optical isolators.  We check whether the non-reciprocal nature is common to all $\cal{PT}$ symmetric systems and, if not, what are the situations under which it can be observed.   Due to the recent interest in studying spontaneous symmetry breaking in coupled waveguide systems, we here present such symmetry breaking bifurcations in certain reciprocal and non-reciprocal $\cal{PT}$ symmetric systems.  
\end{abstract}
\maketitle
\section{Introduction}
 \par Explorations on $\cal{PT}$ symmetric systems \cite{bendr} open up interesting observations and applications in a variety of fields such as optics \cite{r3,r4,r41,r42,r43}, plasmonics \cite{r5,r55}, quantum optics \cite{r6,r7}, Bose-Einstein condensation \cite{r8, r9}, acoustics \cite{acous}, electronics \cite{r11,r12} and mechanics \cite{r13,r14,r15}. In the literature such $\cal{PT}$ symmetric structures were constructed straightforwardly by coupling a system with loss to a system with equal amount of gain.  One can even find interesting experimental realizations of such $\cal{PT}$ symmetric systems \cite{r42,nature}.  However, the latter finds a stringent condition in the practical realization whereby the balance between the loss and gain is expected.   Due to the above, the exploration of $\cal{PT}$ symmetric structures without intrinsic loss-gain is interesting and is more advantageous if one finds a better adaptable structure.\\
\par  In this connection, we here consider the problem of studying the loss-gain free $\cal{PT}$ symmetric systems.  We present one such class of systems in which the $\cal{PT}$ symmetric nature is introduced through the coupling.  In such loss-gain free systems with $\cal{PT}$ symmetric coupling, we study the dynamics in comparison with the usual $\cal{PT}$ symmetric systems with loss-gain and we check whether the dynamics of these systems include the novel characteristics of the usual $\cal{PT}$ symmetric systems with loss-gain.  Due to the existence of loss-gain terms, the usual $\cal{PT}$ symmetric systems are considered as special cases of dissipative systems and are found to lie at the boundary between conservative and dissipative systems. Now considering the systems with $\cal{PT}$ interaction, one can find not only dissipative systems but also conservative systems.  Thus it is also interesting to note the dynamics of conservative $\cal{PT}$ symmetric structures in comparison with the dissipative ones.   \\
\par One of the most interesting and useful characteristics of classical $\cal{PT}$ symmetric systems is the non-reciprocal nature which helps to localize the energy in the gain component of the coupled systems \cite{ref2}.  Thus there occurs unidirectional propagation of light in the usual $\cal{PT}$ symmetric systems. But in the systems which we consider there are no loss or gain terms and thus we look at whether such non-reciprocal nature sustains in these type of $\cal{PT}$ symmetric systems as well.  In other words we search whether the non-reciprocal nature is common to all types of $\cal{PT}$-symmetric systems. If it is not then what are the situations in which one can observe non-reciprocity in these systems?  The present article shows the criterion required for the existence of non-reciprocity in these $\cal{PT}$ symmetric systems is the existence of self-trapping nonlinearity.   Also we note the spontaneous symmetry breaking bifurcations in certain cases of this class of systems. With some conservative cases, we show the spontaneous symmetry breaking through the pitchfork bifurcation in the reciprocal cases and the spontaneous symmetry breaking through a tangent like bifurcation in non-reciprocal cases admitting unidirectional transport of light.\\
\par In order to focus on the above objectives, we present this manuscript in the following form.  In Sec. \ref{model}, we describe the general nature of the models that we intend to consider in the manuscript.  In Sec. \ref{simp}, we consider conservative systems with linear $\cal{PT}$-symmetric coupling. We show that many of these $\cal{PT}$ symmetric conservative structures show reciprocal dynamics.  Thus, we add non-conservative nature into the system in Sec. \ref{lin_n} and study the associated dynamics.  We show the existence of non-reciprocal nature in these non-conservative $\cal{PT}$ symmetric structures and also we discuss the necessary elements such as self-trapping nonlinearity needed for the presence of non-reciprocal nature in the systems with linear-$\cal{PT}$ coupling.  In Sec. \ref{nli}, we investigate systems with nonlinear $\cal{PT}$ coupling and discuss the necessary elements needed for the non-reciprocal nature of the system.  In Sec. \ref{sum}, we present our conclusions.  In Appendix. \ref{ap1}, we give the brief discussion of the analytical results corresponding to a conservative non-reciprocal linear $\cal{PT}$ coupled system. 
\section{\label{model}The general models:}
\par  In the literature, $\cal{PT}$ symmetric systems have been studied intensively with nonlinear Schr\"odinger dimers \cite{ref2, bec, kevre, kivsh, flach, susan,bar1,bar2, bar3,genr, bors} in which a system with loss is coupled to a system with gain as 
\begin{eqnarray}
i \frac{d\phi_1}{dz}&=& i \gamma \phi_1 - k \phi_2+ \alpha G(\phi_1, \phi_2, \phi_1^*, \phi_2^*), \nonumber \\
i \frac{d\phi_2}{dz}&=&-i \gamma \phi_2 - k \phi_1+ \alpha G(\phi_2, \phi_1, \phi_2^*, \phi_1^*).  
\label{gailo}
\end{eqnarray}
\par The above dimer model represents a coupled pair of waveguides in which ${\phi_1(z)}$ and $\phi_2(z)$, respectively, represent the complex amplitudes of light in the two waveguides with respect to the propagation distance $z$. The first term in the right hand side represents the loss-gain term, the second term defines the coupling due to the interaction of evanescent fields and $G(\phi_1, \phi_2, \phi_1^*, \phi_2^*)=-G(-\phi_1, -\phi_2, -\phi_1^*, -\phi_2^*)$ can be any odd order nonlinear term defining the nonlinear interactions with the coupling strength $\alpha$.  Note that the above dimer is also relevant to the description of Bose-Einstein condensates in double well $\cal{PT}$ symmetric potentials \cite{bec}.  The presence of loss-gain terms in (\ref{gailo}) makes the system to be $\cal{PT}$ symmetric, where the $\cal{PT}$ operation is defined by $\phi_1 \rightarrow -\phi_2$, $\phi_2 \rightarrow -\phi_1$, $i \rightarrow -i$ and $z \rightarrow -z$ and that $G(\phi_1, \phi_2, \phi_1^*, \phi_2^*)=-G(-\phi_1, -\phi_2, -\phi_1^*, -\phi_2^*)$.  The dynamics of the above system has been studied extensively in the literature \cite{ref2, kevre, kivsh, flach, susan} with different forms of  $G(\phi_1, \phi_2, \phi_1^*, \phi_2^*)$ \cite{bar1, bar2, bar3,genr} and with nonlinear loss-gain terms \cite{bors}.  
\par In this paper, we consider a class of systems in which the $\cal{PT}$ symmetry is introduced by coupling and not by loss-gain terms.  The systems that we consider are of the form
\begin{eqnarray}
i \frac{d\phi_1}{dz}&=& i a \phi_2 - k \phi_2+ \alpha G(\phi_1, \phi_2, \phi_1^*, \phi_2^*), \nonumber \\
i \frac{d\phi_2}{dz}&=&-i a \phi_1 - k \phi_1+ \alpha G(\phi_2, \phi_1, \phi_2^*, \phi_1^*),  
\label{ccoup}
\end{eqnarray}
which include both linear and nonlinear couplings. Later on in Sec. \ref{nli}, we will also consider fully nonlinear $\cal{PT}$ coupling as well. 
In the above, the first coupling term in the right hand side makes the system to be $\cal{PT}$ symmetric. The simplest situation where one can observe such $\cal{PT}$-symmetric coupling is in the coupled waveguides of magneto-optic materials \cite{circ, yar_p,yariv}.  These magneto-optic materials have the dielectric tensor of the form
\begin{eqnarray}
&{\epsilon}=\epsilon_0\left[\begin{array}{cccc}
\epsilon_x &-i \delta&0  \\
i \delta& \epsilon_y&0 \\
0&0&\epsilon_z
\end{array}\right],
\label{tense}
\end{eqnarray} 
where the off-diagonal elements in (\ref{tense}) help one to achieve the expected $\cal{PT}$ symmetric coupling. Because under phase matching situation, the magneto-optic coupling exist among the modes of the two parallel waveguides which can be described by the coupled-mode equations \cite{yar_p,yariv}
\begin{eqnarray}
i \frac{d \phi_1}{dz}&=&ia \phi_2, \nonumber \\
i \frac{d \phi_2}{dz}&=&-ia \phi_1, 
\label{base}
\end{eqnarray}
where the coupling coefficient $a$ is determined from $\delta$ through Eq. (116) in \cite{yar_p}.  In \cite{yar_p}, the coupling coefficient in a paramagnetic material is derived out as an example and it is found to be $a=VH$, where $V$ is the Verdet constant of the material and $H$ is the applied magnetic field. This clearly shows that the equation in (\ref{base}) is found to be symmetric with respect to the combined $\cal{PT}$ operation $\phi_1 \rightarrow -\phi_2$, $\phi_2 \rightarrow -\phi_1$, $i \rightarrow -i$ and $z \rightarrow -z$.  One can also note in \cite{yar_p} (see Eq. (78)) that the electro-optic coupling can also be written in the same form as in (\ref{base}) under phase matching situation. 
\par In the following sections, we study the dynamics of the systems of the form (\ref{ccoup}) and also we extend our studies to systems with nonlinear $\cal{PT}$ coupling or nonlinear magneto-optic coupling. 
 
 \section{\label{simp}Linear $\cal{PT}$-symmetric coupling: Conservative situations}
\subsection{A simple conservative system with linear $\cal{PT}$-symmetric coupling}
\par  To start with, we consider a simple optical nonlinear $\cal{PT}$-symmetric dimer, 
\begin{eqnarray}
i \frac{d \phi_1}{dz} &=& - \beta |\phi_1|^2 \phi_1 - k \phi_2 + i a \phi_2, \nonumber \\
i \frac{d \phi_2}{dz}  &=&- \beta |\phi_2|^2 \phi_2 - k \phi_1 - i a \phi_1.
\label{lin1}
\end{eqnarray}
\par In the above system, $\beta$ represents the strength of self-focusing nonlinearity, $k$ corresponds to the coupling due to the evanescent fields while $a$ represents the magneto-optic coupling.  In the literature, this type of $\cal{PT}$ symmetric system with the complex coupling has been studied in the presence of loss-gain terms \cite{compact, kev}.  In \cite{compact}, $\cal{PT}$ symmetric compactons are observed in a system of three-line waveguide array where the existence of such compactons requires the presence of loss-gain terms and also the complex coupling.  In \cite{kev}, Stokes variable dynamics of Eq. (\ref{lin1}) with loss-gain terms has been studied as a subcase of a general dimer model.  To our knowledge, the above type of coupled $\cal{PT}$ symmetric system has not been studied in-detail in the absence of loss-gain terms. We here consider such loss-gain free $\cal{PT}$ symmetric models. 
\par One can find that the system (\ref{lin1}) is conservative so that the total power becomes
\begin{eqnarray}
&\frac{dP}{dz}=0, \qquad \qquad \qquad& 
\end{eqnarray}
where
\begin{eqnarray}
  &P=|\phi_1|^2+|\phi_2|^2. \qquad \qquad \qquad &
\label{pz}
\end{eqnarray}
 In the following, throughout the manuscript, conservative systems are referred to those systems in which the total power is conserved as above.
\par  Now let us look at the light beam propagation with respect to the propagation distance, $z$, and compare its dynamics with that of the usual $\cal{PT}$ symmetric systems with loss-gain profile. For this purpose, we first look into the dynamics in the linear case of the system (\ref{lin1})
\begin{eqnarray}
i \frac{d \phi}{dz}={\cal{H}} \phi,
\end{eqnarray}
where $\phi=[\phi_1  \;\;  \phi_2]^T$ and the Hamiltonian $\cal{H}$
\begin{eqnarray}
&{\cal{H}}=\left(\begin{array}{cccc}
0 &-k+ia  \\
 -k-ia&0 \\
\end{array}\right). 
\end{eqnarray}  
The corresponding eigenvalues of the linear problem are $\lambda = \pm \sqrt{a^2+k^2}$.  This shows the existence of a completely real eigenvalue spectrum which corresponds to unbroken $\cal{PT}$ symmetry for all parametric values of ($k,a$).  This is in contrast to the linear $\cal{PT}$ symmetric system with loss-gain (Eq. (\ref{gailo}) without nonlinear terms) where the eigenvalues are $\lambda=\pm \sqrt{k^2-\gamma^2}$ \cite{ref2}.  Here, an increase in loss-gain strength makes the eigenvalues to be imaginary and thus the symmetry will be spontaneously broken.  
\begin{figure}[]
\hspace{-1.0cm}
   \includegraphics[width=1.11\linewidth]{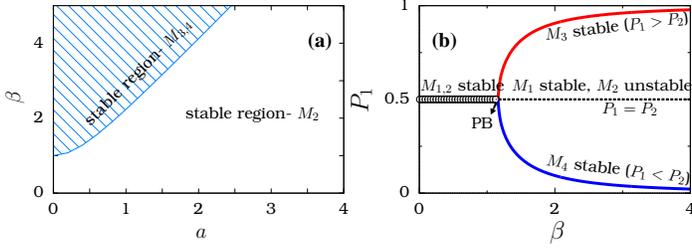}
   \caption{(color online) (a) Figure shows the stable regions of symmetric as well as asymmetric fixed points of Eq. (\ref{lin1}) where we have fixed $P(0)=P=1.0$, $k=0.5$.  Note that in the entire region, $M_1$ is stable. Fig. (b) shows the pitchfork bifurcation for $k=0.5$, $P=1.0$ and $a=0.3$. }
\label{f0}   
\end{figure} 
\par With the introduction of the self-trapping nonlinearity, we now check whether the nonlinear modes of light can break the symmetry spontaneously.  For this purpose, we consider $\phi_1(z)= R_1(z) e^{i (\omega z+\theta_1)}$, $\phi_2(z) =R_2(z) e^{i (\omega z+\theta_2)}$ in Eq.(\ref{lin1}), where $R_i(z),$ $i=1,2$ are the wave amplitudes, $\omega$ is the propagation constant and $\theta_i$, $i=1,2$ are the constant phases.  Upon substitution, one will obtain the dynamical equations corresponding to the amplitude and the phase difference $\delta=\theta_1-\theta_2$ as
\begin{eqnarray}
\dot{R_1}&=&k R_2 \sin \delta+a R_2 \cos \delta, \label{e1_1} \\
\dot{R_2}&=&-k R_1 \sin \delta-a R_1 \cos \delta,  \label{e2_1} \\
\dot{\delta}&=&(\beta R_1 R_2-k \cos \delta+a \sin \delta)\frac{(R_1^2-R_2^2)}{R_1 R_2}. 
\label{e3_1}
\end{eqnarray}
Note that the above equations are equivalent to the system (\ref{lin1}).  As the system is conservative, we reduce the order of the above equations (\ref{e1_1})-(\ref{e3_1}) by taking $R_2=\sqrt{P-R_1^2}$.  Thus the defining equations will reduce to the form
\begin{eqnarray}
&\dot{R_1}=(k  \sin \delta+ a \cos \delta)\sqrt{P-R_1^2},& \label{sec1_1} \\
&\dot{\delta}=(\beta R_1 \sqrt{P- R_1^2}-k \cos \delta+a \sin \delta)\frac{2 R_1^2-P}{R_1 \sqrt{P-R_1^2}},&
\label{sec2_1}
\end{eqnarray}
  The stationary nonlinear modes correspond to the situation $\dot{R_1}=\dot{\delta}=0$.   Thus from Eqs. (\ref{sec1_1}-\ref{sec2_1}), one can find the existence of symmetric nonlinear modes of the form
\begin{eqnarray}
M_1:\;R_1 \equiv R_1^*&=\sqrt{P/2}, \quad \cos \delta^*=\frac{-k}{\sqrt{a^2+k^2}}, \label{mm1}\\
M_2: \;R_1 \equiv R_1^*&=\sqrt{P/2}, \quad \cos \delta^*=\frac{k}{\sqrt{a^2+k^2}}, \label{mm2}
\end{eqnarray}
where $\sin \delta^*=\frac{-a \cos \delta^*}{k}$.  The modes are said to be symmetric as $R_1^*=R_2^*$, where $R_2^*=\sqrt{P-{R_1^*}^2}$.  Note that in Eq. (\ref{sec2_1}), $R_1^*=P$ leads to a singularity.   One can also find the existence of asymmetric nonlinear stationary modes (where $R_1^* \neq R_2^*$) of the form
\begin{eqnarray}
M_{3}:\;R_1 \equiv R_1^*&=\sqrt{\frac{P \beta + \sqrt{P^2 \beta^2-4 (a^2+k^2)}}{2 \beta}},  \label{mm3} \\
M_{4}: \;R_1 \equiv R_1^*&=\sqrt{\frac{P \beta - \sqrt{P^2 \beta^2-4 (a^2+k^2)}}{2 \beta}}. \label{mm4} 
\end{eqnarray}
For both modes $M_3$ and $M_4$, $\sin \delta^*= \frac{-a}{\sqrt{a^2+k^2}}$ and $\cos \delta^*=\frac{k}{\sqrt{a^2+k^2}}$.  Also from (\ref{mm3}) and (\ref{mm4}), we can find that these asymmetric modes exist only when $\beta \geq \frac{ 2\sqrt{a^2+k^2}}{P}$. 
\par The linear stability of each symmetric and asymmetric modes has been studied and the results show that the eigenvalues corresponding to each mode are
\begin{eqnarray}
M_1: \; \lambda&=& \pm i 2\sqrt{\sqrt{a^2+k^2} \left(\frac{\beta P}{2}+\sqrt{a^2+k^2}\right)}, \quad \label{lm1} \\
M_2: \; \lambda&=& \pm 2\sqrt{\sqrt{a^2+k^2} \left(\frac{\beta P}{2}-\sqrt{a^2+k^2}\right)},\quad \label{lm2} \\
M_3: \; \lambda&= &\pm i \sqrt{P^2 \beta^2 - 4(a^2+k^2)}, \quad \label{lm3} \\
M_4: \; \lambda&=& \pm i \sqrt{P^2 \beta^2 - 4(a^2+k^2)},. \label{lm4}
\end{eqnarray}
From the eigenvalues corresponding to the symmetric modes $M_1$ and $M_2$ given in Eqs. (\ref{lm1}) and (\ref{lm2}), we can find that the mode $M_1$ is found to be neutrally stable for all parametric values whereas the mode $M_2$ is neutrally stable only when $\beta<\frac{ 2\sqrt{a^2+k^2}}{P}$.  In the above, neutral stability corresponds to the pure imaginary eigenvalues.   The region in which this symmetric mode $M_2$ is stable is also shown in Fig. \ref{f0}(a).  From Eqs. (\ref{lm3}) and (\ref{lm4}), one can find that the eigenvalues corresponding to $M_3$ and $M_4$ are the same and these asymmetric modes are found to be neutrally stable when $\beta >\frac{ 2\sqrt{a^2+k^2}}{P}.$   The stable region of these asymmetric modes is also shown in Fig. \ref{f0}(a).  Note that in the region $\beta >\frac{ 2\sqrt{a^2+k^2}}{P}$, not only the asymmetric modes are stable but also the symmetric mode $M_1$ is found to be  neutrally stable.
\par From the above, one can note that at $\beta=\frac{2 \sqrt{a^2+k^2}}{P}$, a pitchfork like bifurcation occurs which stabilizes the two asymmetric modes $M_3$ and $M_4$ and breaks the $\cal{PT}$ symmetry spontaneously (since $R_1^* \neq R_2^*$).  Such pitchfork bifurcation (PB) in the system is illustrated in Fig. \ref{f0}(b).  The figure shows that after the pitchfork bifurcation point, the modes $M_1$, $M_3$ and $M_4$ are all stable.  Thus depending upon the input power distribution any of the modes $M_1$, $M_3$ and $M_4$ will appear.  Due to the above fact, if $P_1(0)>P_2(0)$ ($P_2(0)>P_1(0)$), the mode $M_3$ ($M_4$) is stabilized and the light will be localized in the first (second) waveguide and if $P_1(0)$ is near to $P_2(0)$ ($P_1(0) \approx P_2(0)$), the symmetric mode $M_1$ will be stabilized.
\par A couple of recent publications in {\it Nature Photonics} \cite{hamel, malo} deal with the above type of spontaneous symmetry breaking in coupled waveguides where the authors have compared such situations with the spontaneous symmetry breaking in a double well potential through pitchfork bifurcation. They noted that just like the particle in a double well is localized in one of the wells depending upon its initial position, here also depending on the input distribution of light (whether $P_1(0)>P_2(0)$ or $P_2(0)>P_1(0)$) it is localized in one of the waveguides. In the following discussions, we observe that this pitchfork type of symmetry breaking bifurcation has a relationship with the reciprocal nature of the systems. 
\begin{figure}[htb!]
\hspace{-1.0cm}
   \includegraphics[width=1.1\linewidth]{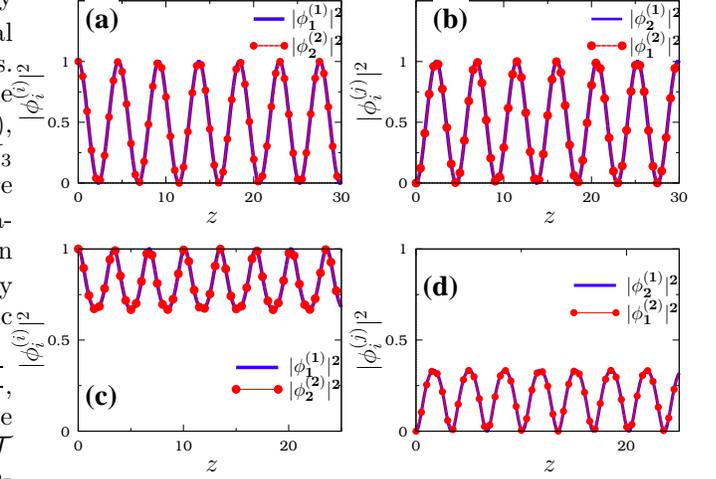}
   \caption{(color online) Reciprocal dynamics in the system (\ref{lin1}) for $k=0.5$, $a=0.5$.  Figs. (a) and (b) show the reciprocal nature of the system in the unbroken case for $\beta=1.0$. From Figs. (a) and (b), it is obvious that $|\phi_1^{(1)}|^2= |\phi_2^{(2)}|^2$ (as both the curves completely overlap with each other) and $|\phi_2^{(1)}|^2= |\phi_1^{(2)}|^2$, thus the system is reciprocal.  In a similar way, Figs. (c) and (d) show the reciprocal nature in the broken $\cal{PT}$ phase of the system (\ref{lin1}) for $\beta=3.0$. }
\label{f1}   
\end{figure}
\begin{figure}[]
\hspace{-1.0cm}
   \includegraphics[width=1.1\linewidth]{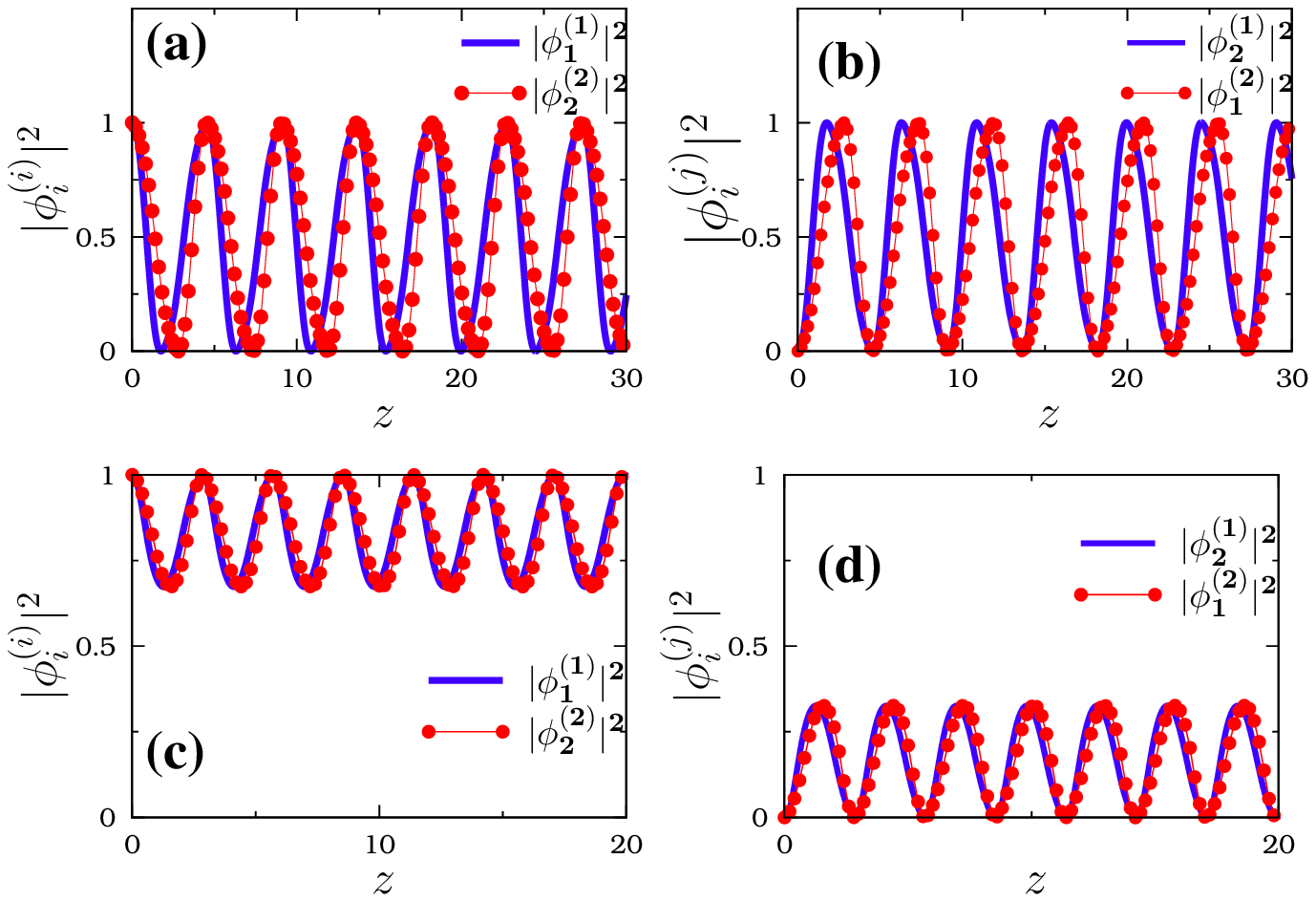}
   \caption{(color online) Dynamics in a conservative case where $G(\phi_1, \phi_2, \phi_1^*, \phi_2^*)= \phi_2^2 \phi_1^*$ in (\ref{gcon}).  Figs. (a) and (b) are plotted in the unbroken $\cal{PT}$ region for $k=0.5$, $a=0.5$, $\beta=2.0$ and $\alpha=1.0$.  Figs. (c) and (d) are plotted in the broken $\cal{PT}$ region for $k=0.5$, $a=0.5$, $\beta=3.5$ and $\alpha=1.0$. By looking at the Figs. (c) and (d) and comparing it with Figs. \ref{f1}(c) and \ref{f1}(d), it is obvious that a pitchfork like symmetry breaking bifurcation occurs in the system. But here the system is not completely reciprocal, as $|\phi_1^{(1)}|^2 \neq |\phi_2^{(2)}|^2$ (see Figs. (a) and (c)) and $|\phi_2^{(1)}|^2 \neq |\phi_1^{(2)}|^2$ (see Figs. (b) and (d)).  We observe phase differences among ($|\phi_1^{(1)}|^2,|\phi_2^{(2)}|^2$) and ($|\phi_2^{(1)}|^2 ,|\phi_1^{(2)}|^2$). }
\label{ss1}   
\end{figure}
\subsubsection{Reciprocal nature}
\par The reciprocal nature of the system represents the time reversibility and the non-reciprocal nature of the system arises due to the suppression in the time-reversibility nature. To determine whether the dynamics of the system is reciprocal or non-reciprocal, we consider two input situations: (i) $|\phi_1^{(1)}|^2(0)=1$, $|\phi_2^{(1)}|^2(0)=0$ (ii) $|\phi_1^{(2)}|^2(0)=0$, $|\phi_2^{(2)}|^2(0)=1$.  If the system is reciprocal, in both the unbroken and broken $\cal{PT}$ regions one can observe exact matching in the beam propagation patterns in the form $|\phi_1^{(i)}(z)|^2=|\phi_2^{(j)}(z)|^2$, where $i,j \in 1,2$ and $i \neq j$. If the systems shows non-reciprocal dynamics, we find $|\phi_1^{(i)}(z)|^2\neq|\phi_2^{(j)}(z)|^2$, where $i,j \in 1,2$ and $i \neq j$.   To check the reciprocal nature of the system, we have plotted $|\phi_1^{(1)}|^2$ and $|\phi_2^{(2)}|^2$ in Fig. \ref{f1}(a) and $|\phi_2^{(1)}|^2$ and $|\phi_1^{(2)}|^2$ in Fig. \ref{f1}(b) in the unbroken $\cal{PT}$ region of the system.  The figures clearly illustrate the reciprocal nature of the system, in a similar manner, Figs. \ref{f1}(c) and \ref{f1}(d) are plotted in the broken $\cal{PT}$ region of the system. Again it is clear from both the figures that the system is reciprocal.  
\par Specifically the dynamics in the broken $\cal{PT}$ region (Fig. \ref{f1}(c) and \ref{f1}(d)) emphasizes that the pitchfork bifurcation of asymmetric modes (or the stabilization of both the asymmetric modes (as shown in \ref{f0}(b)), the mode $M_3$ with $P_1^*> P_2^*$ and $M_4$ with $P_2^*>P_1^*$, where $P_1^*$ and $P_2^*$ are the corresponding powers) is necessary for the reciprocal nature of the system so that $|\phi_1^{(1)}|^2$ ($|\phi_2^{(1)}|^2$) can be matched with $|\phi_2^{(2)}|^2$ ($|\phi_1^{(2)}|^2$).  This emphasizes the relation between the pitchfork bifurcation and the reciprocal nature of the system.  
\subsection{\label{gcase} General case}
\par  The above discussions clearly show the absence of non-reciprocal nature in the considered simple conservative $\cal{PT}$ symmetric system. To know the presence or absence of such behavior in other conservative $\cal{PT}$ symmetric structures with linear $\cal{PT}$ coupling, we consider a general case
\begin{eqnarray}
i \frac{d \phi_1}{dz} &=-\beta |\phi_1|^2 \phi_1 - k \phi_2 + i a \phi_2+\alpha G(\phi_1, \phi_2, \phi_1^*, \phi_2^*) , \quad \nonumber \\
i \frac{d \phi_2}{dz} &=-\beta |\phi_2|^2 \phi_2 - k \phi_1 - i a \phi_1+ \alpha G(\phi_2, \phi_1, \phi_2^*, \phi_1^*), \quad
\label{gcon}
\end{eqnarray}
where $G(\phi_1, \phi_2, \phi_1^*, \phi_2^*)$ may be any nonlinear term that keeps the system to be conservative and $\cal{PT}$ symmetric.  In the equation corresponding to $\phi_2$, in the function $G$, $\phi_1$ is changed to $\phi_2$ and $\phi_1^*$ is changed to $\phi_2^*$, to maintain $\cal{PT}$ symmetry.  As the coupled mode equations are amplitude equations, $G(\phi_1, \phi_2, \phi_1^*, \phi_2^*)$ is found to be physically acceptable only when the condition \cite{bar2}
\begin{eqnarray}
G(\phi_1 e^{i \chi}, \phi_2 e^{i \chi}, \phi_1^* e^{-i \chi}, \phi_2^* e^{-i \chi})=e^{i \chi}G(\phi_1, \phi_2, \phi_1^*, \phi_2^*).  \quad
\label{connn}
\end{eqnarray}
holds good.  Here $\chi$ is an arbitrary phase constant. 
\begin{figure}[htb!]
\hspace{-1.0cm}
   \includegraphics[width=1.1\linewidth]{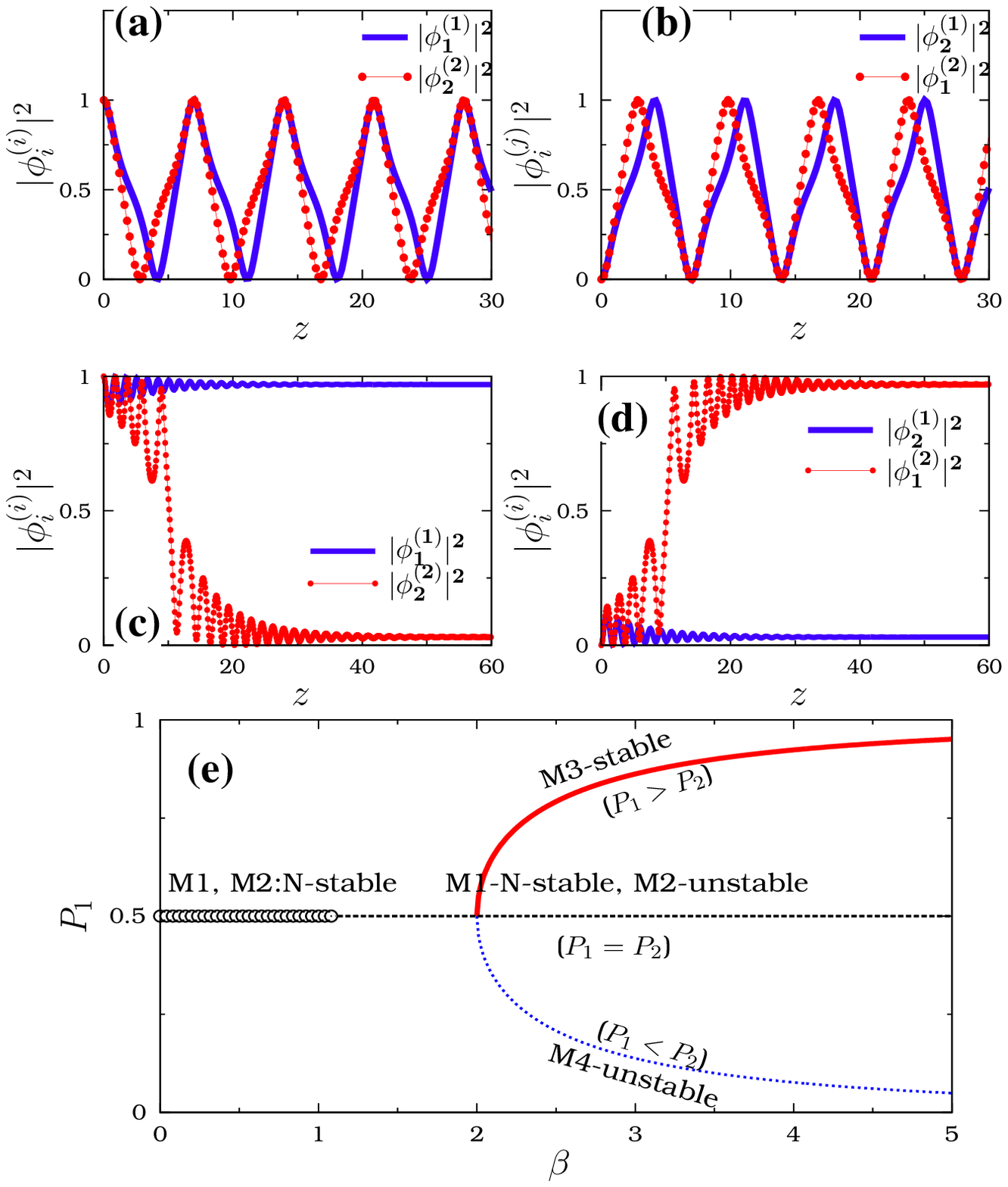}
   \caption{(color online)Non-reciprocal dynamics in the conservative case given in (\ref{gcon}) for $G(\phi_1, \phi_2, \phi_1^*, \phi_2^*)= |\phi_1|^2 |\phi_2^2| \phi_2$.  Figs. (a) and (b) are plotted in the unbroken $\cal{PT}$ region for $k=0.5$, $a=0.5$, $\beta=2.0$ and $\alpha=1.0$.  Figs. (c) and (d) are plotted in the broken $\cal{PT}$ region for $k=0.5$, $a=0.5$, $\beta=4.0$ and $\alpha=1.0$.  The Fig. (e) shows the tangent like bifurcation in the system for $k=0.5$, $a=0.5$, $P=1.0$ and $\alpha=1.0$.}
\label{f12}   
\end{figure}
Thus the possible conservative cubic and quintic nonlinearities are of the forms \cite{bar2}
\begin{eqnarray}
&G(\phi_1, \phi_2, \phi_1^*, \phi_2^*)= |\phi_2|^2 \phi_1,\, \phi_2^2 \phi_1^*,\, \phi_2^3 {\phi_1^*}^2, \, |\phi_1|^4 \phi_1, \, & \nonumber \\  &|\phi_2|^4 \phi_1, \,|\phi_1|^2 |\phi_2^2| \phi_1,\, |\phi_1|^2 |\phi_2^2| \phi_2, \, |\phi_2|^2 \phi_1^2 \phi_2^*. & 
\label{conform}
\end{eqnarray}
\par  While studying the dynamics of the system in the presence of each of the above nonlinear couplings, we observe that in many of the cases except for two quintic nonlinearity forms $G=|\phi_2|^2 \phi_1^2 \phi_2^*, \, |\phi_1|^2 |\phi_2^2| \phi_2$, the symmetry of the system is broken through pitchfork bifurcation as observed in the previous case.  Among the cases showing pitchfork type symmetry breaking, except for the cases with nonlinearities of the form $\phi_2^{n+1} {\phi_1^*}^n$ where $n=1,2$, all the other cases show complete reciprocal dynamics as shown in Fig. \ref{f1}. The nonlinearities $\phi_2^{n+1} {\phi_1^*}^n$ although show similar dynamics as that of Fig. \ref{f1}, there we cannot observe exact matching of $|\phi_1^{(i)}(z)|^2=|\phi_2^{(j)}(z)|^2$, and so they may not be completely reciprocal. The dynamics corresponding to one such cubic nonlinearity case $G=\phi_2^2 \phi_1^*$ has also been shown in Fig. \ref{ss1}.  Due to the pitchfork type symmetry breaking, the foreseen conservative nonlinearities are not useful to achieve unidirectional transport of light as it was achieved in \cite{ref2,r42}.
\par The cases $G=|\phi_2|^2 \phi_1^2 \phi_2^*, \, |\phi_1|^2 |\phi_2^2| \phi_2$ are different from that of the above cases.  These cases are found to be non-reciprocal and are useful for the construction of unidirectional devices.  The dynamics in the case $G=|\phi_1|^2 |\phi_2^2| \phi_2$ is shown in Fig. \ref{f12}.  One can find in Figs. \ref{f12}(a)-\ref{f12}(d) that the dynamics in both the unbroken and broken $\cal{PT}$ phases are found to be non-reciprocal. For this case $G=|\phi_1|^2 |\phi_2^2| \phi_2$, the dynamics of the system has also been studied analytically in appendix \ref{ap1}. The analytical results given in the appendix clearly shows that the asymmetric modes which break the symmetry of the system are found to appear through a tangent like bifurcation and the bifurcation scenario is illustrated with Fig. \ref{f12}(e).  From the above figure, one can note that for small values of $\beta$, the symmetric modes ($M_1$, $M_2$) with $R_1=R_2=P/2=0.5$ are alone found to be stable. Among the symmetric modes $M_1$ is always stable and $M_2$ losses its stability for $\beta >{\frac{2}{P}}\sqrt{a^2+\left(k-\frac{\alpha P^2}{4}\right)^2}$, which has been shown in Fig. \ref{f12}(e).  By increasing the value of $\beta$, two new asymmetric modes ($M_3, M_4$) arise.  Among these asymmetric modes, $M_3$ with $P_1> P_2$ and $M_4$ with $P_1<P_2$, one of them that is $M_3$ alone is stable as it has been shown in Fig. \ref{f12}(e) (Note the other asymmetric modes $M_5$ and $M_6$ do not exist in the considered parametric regions).  This is in contrast to the case with pitchfork bifurcation in which both the asymmetric modes are stable.  Due to the above fact, one can observe localization of light in the first waveguide regardless of the fact that the first waveguide is lighted initially or the second waveguide is lighted initially.  This can be observed clearly if one looks at the curves of $|\phi_1^{(1)}|^2$ and $|\phi_1^{(2)}|^2$ respectively in Figs. \ref{f12}(c) and \ref{f12}(d).   Thus it is clear from the above that there occurs unidirectional transport of light due to the tangent like symmetry breaking bifurcation.
\par From the above discussions on linear $\cal{PT}$ coupled systems, it is obvious that not all conservative $\cal{PT}$ symmetric cases are found to be non-reciprocal but only certain cases are found to be non-reciprocal.  Now by introducing the non-conservative nature into the system itself, we study the dynamics and spontaneous symmetry breaking in the following section.

\section{\label{lin_n}Addition of non-conservative nature}
\par In this section, we consider the non-conservative $\cal{PT}$ symmetrically coupled systems.  We can introduce non-conservative nature into the system through balanced loss-gain terms.  But the presence of non-reciprocal nature in such balanced loss-gain $\cal{PT}$ symmetric system is well known in the literature \cite{ref2, r42}. Thus it is not necessary to study the non-reciprocal nature in the presence of loss-gain terms.  Secondly one of the primary aims of the present work is to find an alternative to loss-gain systems. Thus we look for other types of non-conservative terms such as four wave mixing and find whether such effects can induce non-reciprocal nature in the system. 
\par For this purpose, we consider the system
\begin{eqnarray}
i \frac{d \phi_1}{dz}  &= - \beta |\phi_1|^2 \phi_1 - k \phi_2 + i a \phi_2+ \alpha G(\phi_1, \phi_2, \phi_1^*, \phi_2^*), \quad \nonumber \\
i \frac{d \phi_2}{dz}  &= - \beta |\phi_2|^2 \phi_2 - k \phi_1 - i a \phi_1+ \alpha G(\phi_2, \phi_1, \phi_2^*, \phi_1^*), \quad
\label{lin2}
\end{eqnarray}
 where the functions $G(\phi_1, \phi_2, \phi_1^*, \phi_2^*)$ may correspond to non-conservative nonlinear interaction terms that preserve the $\cal{PT}$ symmetry of the system.  If one considers the nonlinearities upto cubic and quintic orders, one can find the possible physically acceptable nonlinearities satisfying (\ref{connn}) are
\begin{eqnarray}
G(\phi_1, \phi_2, \phi_1^*, \phi_2^*)= \phi_1^2 \phi_2^*,\, |\phi_1|^2 \phi_2,\,|\phi_2|^2 \phi_2, |\phi_1|^4 \phi_2, \nonumber \\ |\phi_2|^4 \phi_2,\, \phi_1^3 {\phi_2^*}^2, \, |\phi_1|^2 \phi_1^2 \phi_2,\, |\phi_1|^2 \phi_2^2 \phi_1^*, \, |\phi_2|^2 \phi_2^2 \phi_1^*.
\label{nnn1}
\end{eqnarray}
  The above expressions contain all possible forms of $G$ upto quintic order which are non-conservative, that is $\frac{dP}{dz} \neq 0$.  One can find nice discussions on the relevance of the above type nonlinearities of cubic order to the physical situations in Barashenkov et al \cite{bar2}.  While studying the dynamics of the system with the above forms of nonlinearities (\ref{nnn1}), we find that all of them are non-reciprocal too.  
\par To illustrate the above, we consider the nonlinearities of cubic order and present their dynamics in detail.  We consider $G$ in (\ref{lin2}) as
\begin{eqnarray}
\alpha G({\phi_1, \phi_2, \phi_1^*, \phi_2^*})= \alpha_1 \phi_1^2 \phi_2^*+\alpha_2 |\phi_1|^2 \phi_2 + \alpha_3 |\phi_2|^2 \phi_2. \quad  
\label{gg}
\end{eqnarray}
One can find that the above nonlinearities correspond to non-conservative situations as 
\begin{eqnarray}
\frac{dP}{dz}=(-\alpha_1+\alpha_2-\alpha_3)S_2 (P_1-P_2) \neq 0.  
\label{dp1}
\end{eqnarray}
The conservative situation arises only when $\alpha_1+\alpha_3=\alpha_2$.  In the above, $S_2$ is one of the Stokes' variables ($S_2=i (\phi_1 \phi_2^*-\phi_1^* \phi_2)$), $P_1$ ($=|\phi_1|^2$) and $P_2$ ($=|\phi_2|^2$) are the powers in the first and second waveguides. 
By introducing the terms of Eq. (\ref{gg}) one by one, we study the dynamics of the system (\ref{lin2}) in the following subsections. 

\subsection{\label{sys1} Case: $\alpha_1 \neq 0$, $\alpha_2=\alpha_3=0$ }
\par First we study the dynamics of the system (\ref{lin2}) in the presence of the four wave mixing term alone corresponding to $\alpha_1 \neq 0$ and $\alpha_{2}= \alpha_3=0$.   As we did in Sec. \ref{simp}, we substitute $\phi_1(z)= R_1 e^{i (\omega z+\theta_1)}$ and $\phi_2(z)= R_2 e^{i (\omega z+\theta_2)}$ and find the equations corresponding to the amplitude and the phase difference which are given by 
\begin{eqnarray}
&\dot{R_1}=(k+\alpha_1 R_1^2) R_2 \sin \delta+ a R_2 \cos \delta,& \label{e1_2} \\
&\dot{R_2}= -(k+\alpha_1 R_2^2) R_1 \sin \delta- a R_1 \cos \delta,& \label{e2_2} \\
&\hspace{-0.3cm}\dot{\delta}=(\beta R_1 R_2- k \cos \delta+a \sin \delta)\frac{R_1^2 -R_2^2}{R_1 R_2}, \; \delta=\theta_1-\theta_2.& \qquad \label{e3_2}
\end{eqnarray}
The system in (\ref{lin2}) with $\alpha_1 \neq 0$, $\alpha_2=\alpha_3=0$ has the following constant of motion
\begin{eqnarray}
C=\beta |\phi_1|^2 |\phi_2|^2- i a (\phi_1 \phi_2^*-\phi_1^* \phi_2)-k(\phi_1 \phi_2^*+\phi_1^* \phi_2). \quad
\label{inte}
\end{eqnarray}
Consequently we have 
\begin{eqnarray}
C= \beta R_1^2 R_2^2- 2 R_1 R_2 (k \cos \delta -a \sin \delta),
\label{r1c}
\end{eqnarray}
so that we can replace $R_2$ in  Eqs. (\ref{e1_2})-(\ref{e3_2}) by
\begin{eqnarray}
R_2= \frac{f(\delta) \pm \sqrt{f^2(\delta)+C \beta}}{\beta R_1},
\label{r2c}
\end{eqnarray}
where $f(\delta)=k \cos \delta- a \sin \delta$.
\par Choosing $R_2=\frac{f(\delta) + \sqrt{f^2(\delta)+C \beta}}{\beta R_1}$ and substituting it in Eqs. (\ref{e1_2})-(\ref{e3_2}), we obtain
\begin{small}
\begin{eqnarray}
&\dot{R_1}=\left((k+\alpha_1 R_1^2) \sin \delta +a \cos \delta \right)\frac{\left(f(\delta)+\sqrt{f(\delta)^2+\beta C}\right)}{\beta R_1},& \label{e1_22} \\
&\dot{\delta}=\frac{\sqrt{f(\delta)^2 + C\beta}\left(\beta^2 R_1^4 -(f(\delta)+\sqrt{f(\delta)^2+C \beta})^2\right)}{\beta R_1^2(f(\delta+\sqrt{f(\delta)^2+C \beta})}.& \label{e3_22} 
\end{eqnarray}
\end{small}
When we try to find out the nonlinear modes of the system, one can find that the modes satisfy
\begin{small}
\begin{eqnarray}
&\cos \delta^*= \pm \frac{(k+\alpha_1 {R_1^*}^2)}{\sqrt{a^2+(k+\alpha_1 {R_1^*}^2)^2}}, \; \sin \delta^*=-\frac{a \cos \delta^*}{(k+\alpha_1 {R_1^*}^2)}, \quad \label{de_st} \\
&\beta^2 {R_1^*}^4-(f(\delta^*)+\sqrt{f(\delta^*)^2+C \beta})^2=0. \label{r1_st}
\end{eqnarray}
\end{small}
\begin{figure}[ht!]
\begin{center}
   \includegraphics[width=1.1\linewidth]{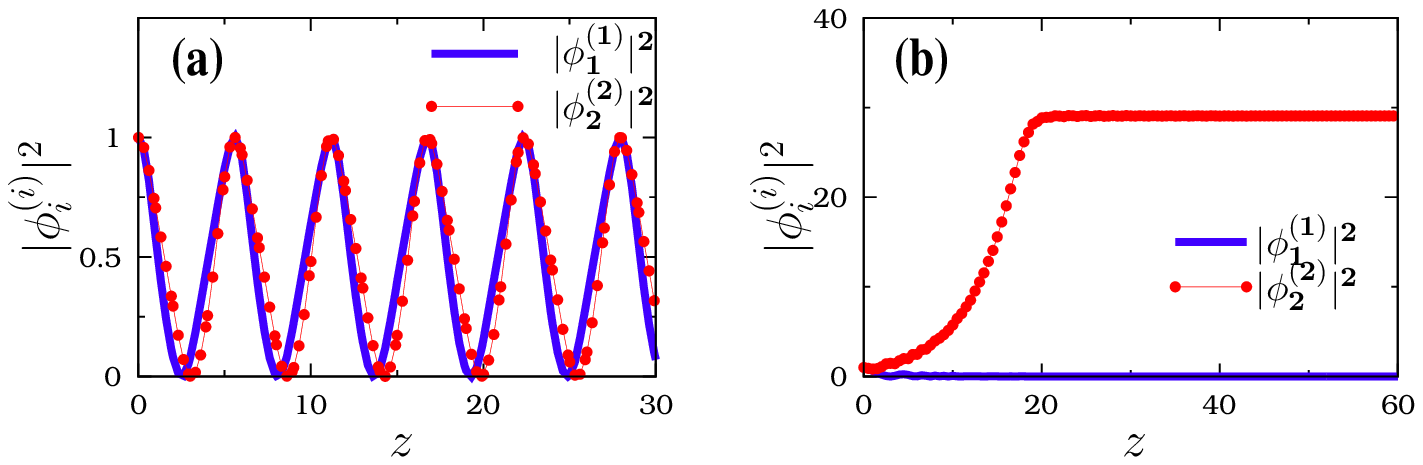}
\end{center}
   \caption{(color online) Figs. (a) and (b) are plotted in the unbroken and broken regions respectively to show the non-reciprocal nature of the system (\ref{lin2}) with $\alpha G=\alpha_1 \phi_1^2 \phi_2^*$.  Here, we obtained the figures by numerically solving Eq. (\ref{lin2}) for the values $a=0.5$, $k=0.3$, $\alpha_1=0.5$ and $\beta=1.5$ and $2.5$, respectively, for Figs. (a) and (b).  }
\label{f3}   
\end{figure}
Note that Eq. (\ref{r1_st}) yields $R_1^*=\sqrt{\frac{(f(\delta^*)+\sqrt{f(\delta^*)^2+C \beta})}{\beta}}$.  By substituting it in (\ref{r2c}), we find that both the modes are symmetric, where, $R_1^*=R_2^*$.   On analyzing the stability of the obtained modes, we find that the modes corresponding to $\cos \delta^*=  \frac{(k+\alpha_1 {R_1^*}^2)}{\sqrt{a^2+(k+\alpha_1 {R_1^*}^2)^2}}$ have the eigenvalues
\begin{small}
\begin{align}
\hspace{-0.1cm}\lambda=\pm\left(\frac{4 a^2 \alpha_1^2 {R_1^*}^4+4 (a^2+(k+\alpha_1 {R_1^*}^2)^2)^{\frac{3}{2}} \sqrt{f^2(\delta^*)+C \beta}}{a^2+(k+\alpha_1 {R_1^*}^2)^2}\right)^{\frac{1}{2}}, \nonumber \\
\label{unst}
\end{align}
\end{small}
where
\begin{small}
\begin{align} 
\sqrt{f^2(\delta^*)+C \beta}=\frac{1}{\sqrt{a^2+(k +\alpha_1 {R_1^*}^2)}} \times \qquad \qquad \qquad \quad  \nonumber \\  \quad \left(\beta {R_1^*}^2 \sqrt{a^2+(k +\alpha_1 {R_1^*}^2)}+(a^2+k(k+\alpha_1 {R_1^*}^2))\right).
\label{fdel}
\end{align}
\end{small}
In the above $R_1^*$ is a solution of  (\ref{r1_st}).
 As both the terms in Eq. (\ref{unst}) are greater than zero, all the modes corresponding to this case are unstable. 
\par For the modes with $\cos \delta^*=  \frac{-(k+\alpha_1 {R_1^*}^2)}{\sqrt{a^2+(k+\alpha_1 {R_1^*}^2)^2}}$, the eigenvalues are
\begin{small}
\begin{eqnarray}
\lambda=\pm i \sqrt{\frac{-4 a^2 \alpha_1^2 {R_1^*}^4+4 (a^2+(k+\alpha_1 {R_1^*}^2)^2)^{\frac{3}{2}} \sqrt{f^2(\delta)+C \beta}}{a^2+(k+\alpha_1 {R_1^*}^2)^2}}. \nonumber \\
\label{nst}
\end{eqnarray}
\end{small}
where the value of $\sqrt{f^2(\delta^*)+C \beta}$ is given in (\ref{fdel}). Looking at the eigenvalues given in Eq. (\ref{nst}), we find that the second term inside the square root is strongly positive, so that the eigenvalues will be pure imaginary. The symmetric modes corresponding to this case are found to be neutrally stable. 
\par From (\ref{r2c}), we can note that $R_2$ can also take the form $R_2=\frac{f(\delta) - \sqrt{f^2(\delta)+C \beta}}{\beta R_1}$, so that Eqs. (\ref{e1_2})-(\ref{e3_2}) can be written as
\begin{small}
\begin{eqnarray}
&\dot{R_1}=\left((k+\alpha_1 R_1^2) \sin \delta +a \cos \delta \right)\frac{\left(f(\delta)-\sqrt{f(\delta)^2+\beta C}\right)}{\beta R_1},& \quad \label{e1_22} \\
&\dot{\delta}=\frac{-\sqrt{f(\delta)^2 + C\beta}\left(\beta^2 R_1^4 -(f(\delta)-\sqrt{f(\delta)^2+C \beta})^2\right)}{\beta R_1^2(f(\delta-\sqrt{f(\delta)^2+C \beta})}.& \label{e3_22} \quad
\end{eqnarray}
\end{small}
In this case, one can find that the stationary nonlinear modes satisfy
\begin{small}
\begin{eqnarray}
&\cos \delta^*= \pm \frac{(k+\alpha_1 {R_1^*}^2)}{\sqrt{a^2+(k+\alpha_1 {R_1^*}^2)^2}}, \;\; \sin \delta^*=-\frac{a \cos \delta^*}{(k+\alpha_1 {R_1^*}^2)},& \label{de_st} \\
&\beta^2 {R_1^*}^4-(f(\delta^*)-\sqrt{f(\delta^*)^2+C \beta})^2=0.& \label{r1_st2}
\end{eqnarray}
\end{small}
The eigenvalues corresponding to these cases are found to be the same as that of the previous case. If $\cos \delta^*=  \frac{(k+\alpha_1 {R_1^*}^2)}{\sqrt{a^2+(k+\alpha_1 {R_1^*}^2)^2}}$, we find that the eigenvalues are the same as the ones given in (\ref{unst}) and if $\cos \delta^*= -\frac{(k+\alpha_1 {R_1^*}^2)}{\sqrt{a^2+(k+\alpha_1 {R_1^*}^2)^2}}$, we find that the expressions for the eigenvalues match with the ones given in (\ref{nst}).  
\par The above results imply that the symmetric modes are neutrally stable for almost all parametric values.   Although the asymmetric modes have not been deduced in the above, the numerical results confirm the existence of such solutions.  However in the asymmetric mode case, the frequency of $\phi_1$ and $\phi_2$ may be different, and so the above analytical results fail to show their existence. 
\begin{figure}[]
   \includegraphics[width=1.05\linewidth]{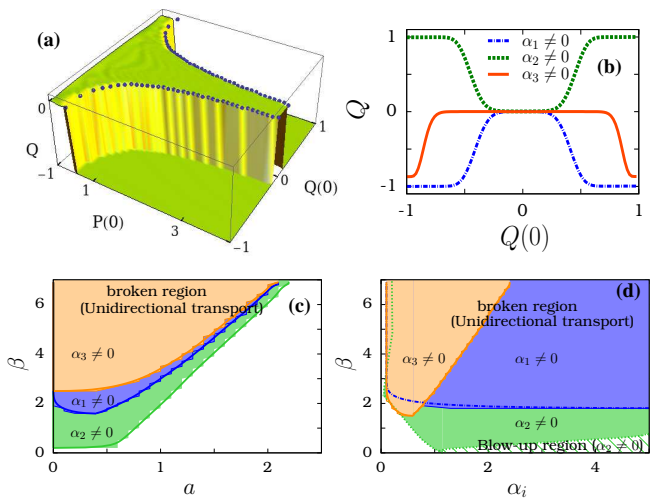}
   \caption{(color online) Fig. (a) The plot of normalized power difference $Q$ for different $Q(0)$ and $P(0)$ in the case $\alpha_1 \neq 0$ and $\alpha_2=\alpha_3=0$, for $\beta=1.0$, $\alpha_1 =0$, $a=0.5$ and $k=0.3$. Fig. (b) shows the variation of $Q$ with respect to $Q(0)$ in all the three cases (i) $\alpha_1 \neq 0$, (ii) $\alpha_2 \neq 0$, (iii) $\alpha_3 \neq 0$ for $a=0.5$, $k=0.3$, $\beta=4.0$ and $\alpha_i=1.0$. Figs. (c) and (d) show the broken $\cal{PT}$ regions in which the light is transported unidirectionally in ($a, \beta$) and ($\alpha_i, \beta$) parametric spaces where Fig. (c)  is plotted for $k=0.3$ and $\alpha_1=1.0$ and for Fig. (d) $k=0.3$ and $a=0.5$. }
\label{f4}   
\end{figure}
\subsubsection{Non-reciprocal nature}
\par   Now we consider the reciprocal or non-reciprocal nature of the system (\ref{lin2}) with $\alpha_1 \neq 0$, $\alpha_2=\alpha_3=0$.  As we did in the earlier section, we here plot $|\phi_1^{(1)}(z)|^2$ and $|\phi_2^{(2)}(z)|^2$ in both the unbroken and broken $\cal{PT}$ phases of the system in Figs. \ref{f3}(a) and \ref{f3}(b).  The asymmetric propagation of light shown in Fig. \ref{f3}(b) has been obtained by solving the system numerically.  From the Figs. \ref{f3}(a) and \ref{f3}(b), it is obvious that $|\phi_1^{(i)}(z)|^2 \neq |\phi_2^{(j)}(z)|^2$, where $i,j \in 1,2$ and $i \neq j$.  Thus this non-conservative case is found to be non-reciprocal.   Also one can find that the light is localized in the second waveguide.
\par Knowing that the system (\ref{lin2}) exhibits non-reciprocal nature, we now look at how far it can isolate light in a waveguide.  For this purpose, we now consider the normalized power difference given by
\begin{eqnarray}
Q= \frac{\overline{P_1}-\overline{P_2}}{\overline{P_1}+\overline{P_2}},
\label{qform}
\end{eqnarray}
where $\overline{P_1}$ ($\overline{P_2}$) represents the average of $P_1$ ($P_2$) with respect to $z$.  Here, $P_1$ and $P_2$ are the quantities $P_1=|\phi_1|^2$ and $P_2=|\phi_2|^2$.  If $Q$ is zero then the light is equally distributed in the first and second waveguides so that there occurs propagation of symmetric modes in the system.  Whenever $Q$ takes the values $1$ or $-1$, we can find that the light is well localized in one of the waveguides, namely the first (while $Q=1$) or second (while $Q=-1$) waveguide.  If $Q$ takes up intermediate values ($-1<Q<0$ and $0<Q<1$) then there exists an asymmetric state in which light is not localized completely in either one of the waveguides. 
\par By fixing the parameters as $a=0.5$, $k=0.3$, $\alpha_1=1.0$ and $\beta=2.5$ (which lie in the broken $\cal{PT}$ region as checked numerically), in Fig. \ref{f4}(a) we plot the normalized input power difference $Q$ with different input powers ($P(0)$) and also for different input power distributions $(Q(0))$, where the input power distributions are characterized by 
\begin{eqnarray}
Q(0)= \frac{{P_1}(0)-{P_2}(0)}{{P_1}(0)+{P_2}(0)}.
\label{q0form}
\end{eqnarray}
Here, $P_1(0)$ and $P_2(0)$ represent the power given to the first and second waveguides initially and $P_1(0)+P_2(0)=P(0)$.  Note that $Q(0)=0$ represents the symmetric input power distribution, while non-zero values of $Q(0)$ represent the asymmetric distribution of input power among the waveguides.  From Fig. \ref{f4}(a), one can find that when $P(0)$ is small, the system remains in the symmetric state (as $Q=0$) for all values of $Q(0)$. Increasing the total input power $P(0)$, one can find in Fig. \ref{f4}(a) that $Q$ becomes zero only when the input power distribution is also nearly symmetric (or near $Q(0)=0$)and if the asymmetry in the input power distribution is increased (that is, if $Q(0)$ is away from $0$), the value of $Q$ sharply jumps to $-1$.  In the latter case, we can also note in Fig. \ref{f4}(a) that the value of $Q$ is $-1$ at both the extremes of $Q(0)$ representing the non-reciprocal nature of the system. This shows that the system can be well used for unidirectional transport of light.  
\par However, for low input powers, the asymmetric modes are not stable and in this region the linear modes come into play.  As these linear modes are stable for all parametric ranges, isolation of light cannot be achieved with low input powers (as can be seen in Fig. \ref{f4}(a)).  This implies that the above type of systems cannot be used for low power applications.  On the other hand in the $\cal{PT}$ symmetric system with linear loss-gain, the linear modes lose their stability with an increase of loss-gain strength (as shown in Sec. \ref{simp}).  Thus the $\cal{PT}$ symmetric systems with loss-gain may have an advantage during low power applications, though balancing loss and gain in such systems may be an issue. 
\subsection{\label{other} Other Cases:}
\par We have also verified the existence of non-reciprocal nature in the presence of other non-conservative terms in Eq. (\ref{gg}) individually, that is we considered the cases $\alpha_2 \neq 0$, $\alpha_1=\alpha_3=0$ and $\alpha_3 \neq 0$, $\alpha_1=\alpha_2=0$.  In both these cases, we find the existence of non-reciprocal nature in the system. Thus one finds that the non-conservative nature can induce non-reciprocal nature in the systems with linear $\cal{PT}$ coupling. In Fig. \ref{f4}(b), we have shown how far the above discussed cases (i) $\alpha_1 \neq 0$, (also $\alpha_2=\alpha_3=0$), (ii) $\alpha_2 \neq 0$, (also $\alpha_1=\alpha_3=0$) and (iii) $\alpha_3 \neq 0$, (also $\alpha_1=\alpha_2=0$) are found to be useful in optical isolation.  For these cases, we have plotted the values of $Q$ for different $Q(0)$, where $Q$ and $Q(0)$ are given respectively in Eqs. (\ref{qform}) and (\ref{q0form}).  From Fig. \ref{f4}(b), one can observe that in all the three cases around $Q(0)=0$, the system remains in the symmetric state where $Q=0$.  When the value of $Q(0)$ departs more from $0$, we find the appearance of the asymmetric state corresponding $Q \neq 0$.  Fig. \ref{f4}(b) also reveals that in the first and third cases mentioned above, we can note that the asymmetric state is characterized by $Q=-1$. In the second case corresponding to $\alpha_2 \neq 0$, $Q$ reaches $1$. This denotes that in the first and third cases, the light is isolated in the second waveguide whereas in the second case, the light is isolated in the first waveguide. From Fig. \ref{f4}(b) one can also note that $Q$ does not reach $-1$ fully in the third case, which denotes that the system weakly localizes the light in the second waveguide. Thus one may need a higher value of $\beta$ in order to achieve perfect isolation in the third case. 
\par Although in the above discussions, we demonstrated the dynamics by introducing one by one the terms given in (\ref{gg}), we can also take the combination of these terms.  From Fig. \ref{f4}(b) we can note that the terms corresponding to $\alpha_1$ and $\alpha_3$ have the capability to localize light in the second waveguide and the term corresponding to $\alpha_2$ has the capability to localize light in the first waveguide. Thus when $\alpha_1$ and $\alpha_3$ alone are set to be non-zero, the localization of light occurs in the second waveguide.  If $\alpha_2$ is also made non-zero with the other terms, depending upon whether $\alpha_2$ is greater than or lower than the other quantities $\alpha_1$ and $\alpha_3$, localization in the first or second waveguide occurs. Note that, in the situations (a) $\alpha_1 = \alpha_2$, $\alpha_3=0$, (b) $\alpha_2=\alpha_3$, $\alpha_1=0$ and (c) $\alpha_1+\alpha_3= \alpha_2$, the system becomes conservative where the change in the total power of the system (\ref{lin2}) is given by Eq. (\ref{dp1}).   In these conservative cases, the system shows reciprocal dynamics. 
\par In Figs. \ref{f4}(c) and \ref{f4}(d), we note the parametric regions in which the unidirectional transport of light can be achieved for the three particular cases mentioned earlier by (i)-(iii).  From Fig. \ref{f4}(c), we note that the unidirectional transport occurs only when the $\cal{PT}$ coupling strength is low (but $a \neq 0$) and that the strength of nonlinearity is sufficiently large.   In Fig. \ref{f4}(d), we have plotted such unidirectional transport regions in ($\beta$-$\alpha_i$) space.  From both the figures we can note that such unidirectional region corresponding to the third case (iii) $\alpha_3 \neq 0$, $\alpha_1=\alpha_2=0$ is not wider compared to the other cases.     From both these figures we can also note that the parameter $\beta$ plays an important role in inducing unidirectional transport of light. Even the increase in $\alpha_i$'s is not contributing much to the emergence of unidirectional propagation region.  Also one may find that at $\beta=0$, the unidirectional transport of light is absent.  The lack of localizing nature of nonlinear interaction terms $\alpha_i$'s may be the reason for the absence of such dynamics at $\beta=0$. 
\par  In this connection, we note the necessity of the self-trapping nonlinearity for the unidirectional transport of light.  We expect that the absence of self-trapping nonlinearity only suppresses the localization of light in one of the waveguides.  However, the dynamics of the system is expected to be non-reciprocal. But our observation shows that the absence of self-trapping nonlinearity not only impacts the localization of light but also the non-reciprocal nature of the system.  More clearly, in the case $\beta=0$, we expect the beam propagation in the waveguide follows the form shown in Fig. \ref{f3}(a) (where we have pointed out the non-reciprocal propagation of light when the symmetry is unbroken).   But the system actually exhibits a reciprocal propagation as in Fig. \ref{f1}(a) or \ref{f1}(b).   Thus it is clear from the discussion of nonconservative cubic nonlinearities that the presence of self-trapping nonlinearity is crucial in inducing non-reciprocal dynamics. 
\par In a similar way, we studied the dynamics in the presence of quintic nonlinearities given in (\ref{nnn1}) and we found that all these non-conservative cases are non-reciprocal. Except for the case $G=|\phi_1|^2 \phi_1^2 \phi_2^*$, all the other cases allow optical isolation in certain parametric regions.  As shown in the cubic nonlinearity cases, in most of the quintic nonlinearity cases (except for the cases $G=|\phi_1|^2 \phi_2^2 \phi_1^*$, $|\phi_2|^2 \phi_2^2 \phi_1^*, \phi_1^3 {\phi_2^*}^2$) self-trapping nonlinearity is found to play a crucial role in inducing non-reciprocity.  For $\beta=0$, one can find that the localization of light could not be achieved in all the cases.  While $\beta=0$, we can observe that the $\cal{PT}$ symmetry is found to be unbroken for certain values of the nonlinearity strength ($\alpha$) (in certain cases, the $\cal{PT}$ symmetry remains unbroken even for higher values of $\alpha$).  We also find that in many cases, increase in the strength of the above nonlinearities lead to blow up solutions.  While looking at the dynamics in the unbroken $\cal{PT}$ region at $\beta=0$, one can note that it is reciprocal for most of the cases.  For few cases like, $G=|\phi_1|^2 \phi_2^2 \phi_1^*$, $|\phi_2|^2 \phi_2^2 \phi_1^*, \phi_1^3 {\phi_2^*}^2$, the dynamics in the unbroken region is non-reciprocal (as similar to the one shown in Fig. \ref{f3}(a)).  As the increase in the strength of such quintic nonlinearities lead to blow up even in the latter cases, we cannot achieve localization of light in a waveguide as it was achieved in Fig. \ref{f3}(b).  Thus it is clear that without the self-trapping nonlinearity, there are no possible application of the above quintic forms of $G$ towards unidirectional devices like optical diodes.

\par Finally, from the discussions given in Secs. \ref{simp} and \ref{lin_n}, one can note the following conclusions on the systems with linear $\cal{PT}$ coupling.  All the non-conservative $\cal{PT}$ symmetric systems are non-reciprocal but only certain conservative cases are non-reciprocal. In view of the above discussions towards the application of optical diodes, we find that the self-trapping nonlinearity is very important.  
\section{\label{nli}Nonlinear $\cal{PT}$ interactions}
\par Now we turn our attention to the cases with nonlinear $\cal{PT}$ interaction,
\begin{eqnarray}
i \frac{d \phi_1}{dz} &=& - \beta |\phi_1|^2 \phi_1 - k \phi_2 + i \eta G(\phi_1, \phi_2, \phi_1^*, \phi_2^*), \nonumber \\
i \frac{d \phi_2}{dz} &=&- \beta |\phi_2|^2 \phi_2 - k \phi_1 - i \eta G(\phi_2, \phi_1, \phi_2^*, \phi_1^*), 
\label{non1}
\end{eqnarray}
where the last term denotes the nonlinear $\cal{PT}$ coupling.   The $i G(\phi_1, \phi_2, \phi_1^*, \phi_2^*)$ term with strength $\eta$ may be any nonlinear term that preserves $\cal{PT}$ symmetry of the system.  Now we first concentrate on the dynamics of nonlinear $\cal{PT}$ coupled systems in the non-conservative situations.  Concerning the physically acceptable forms of $G$ that can make (\ref{non1}) to be non-conservative, we can find 
\begin{eqnarray}
G(\phi_1, \phi_2, \phi_1^*, \phi_2^*)=|\phi_1|^2 \phi_2, |\phi_2|^2 \phi_2, \phi_1^2 \phi_2^*.
\label{gg3non}
\end{eqnarray}
In the above, we presented the forms of $G$ upto cubic order corresponding to the non-conservative cases alone.  The other possible and physically acceptable cubic nonlinearities are conservative and they will be discussed later (see Eq. (\ref{g35con}) below). Studying the dynamics with respect to all the above nonlinearities, we find that the system is non-reciprocal in all the three cases and the unidirectional transport of light can be achieved only in the presence of self-trapping nonlinearity.  As seen in the previous case, the absence of self-trapping nonlinearity impacts the non-reciprocal nature of the system. 
\par If one considers non-conservative $G$ upto quintic order, $G$ can take the forms
\begin{eqnarray}
G(\phi_1, \phi_2, \phi_1^*, \phi_2^*)=|\phi_1|^4 \phi_1, |\phi_1|^4 \phi_2, \, |\phi_2|^4 \phi_1,\, |\phi_2|^4 \phi_2,\, \nonumber \\  |\phi_1|^2 |\phi_2|^2 \phi_1, \, \phi_1^3 {\phi_2^*}^2, \, |\phi_1|^2 \phi_1^2 \phi_2^*,\, |\phi_1|^2 \phi_2^2 \phi_1^*,\, |\phi_2|^2 \phi_2^2 \phi_1^*\quad
\label{gg5non}
\end{eqnarray} 
It has been verified that the system is found to be non-reciprocal in all the above cases.  Many of these quintic nonlinearities are not much advantageous for the application of optical isolation as the $\cal{PT}$ symmetry is unbreakable in some of the cases (cases $G=|\phi_2|^4 \phi_1, |\phi_1|^2 \phi_1^2 \phi_2^*$) and in many of the cases there exists wide blow-up regions and the localization of light has been achieved only with higher values of $\beta$.  In none of the cases, unidirectional transport of light is achieved in the absence of self-trapping nonlinearity.  Again it emphasizes the importance of self-trapping nonlinearity. 
\par These nonlinear $\cal{PT}$ coupled systems also have conservative situations, where $G$ can take the following forms for nonlinearities of order three and five,
\begin{eqnarray}
G(\phi_1, \phi_2, \phi_1^*, \phi_2^*)&=&|\phi_2|^2 \phi_1,\, \phi_2^2 \phi_1^*,\, |\phi_1|^2 |\phi_2|^2 \phi_2, \nonumber \\ && \phi_2^3 {\phi_1^*}^2, |\phi_2|^2 \phi_1^2 \phi_2^*.
\label{g35con}
\end{eqnarray}
Unlike the linear $\cal{PT}$ coupled system, here all the conservative nonlinearities are found to be non-reciprocal, except for the nonlinearities of the form $\phi_2^{n+1} {\phi_1^*}^n$, where $n=1,2$.  As we have seen in Sec. 3.2, the conservative nonlinearities of the form $\phi_2^{n+1} {\phi_1^*}^n$ show non-reciprocal phase shift but they show spontaneous symmetry breaking through the pitchfork bifurcation (the dynamics is similar to that of the one given in Fig. \ref{ss1}).  Thus except for the nonlinearities of the form $\phi_2^{n+1} {\phi_1^*}^n$,  all the other nonlinearities are found to be useful from the perspective of unidirectional devices.  As seen in the previous cases, the self-trapping nonlinearity is found to play an important role in achieving unidirectional devices except for the case of a singular example $G=|\phi_2|^2 \phi_1$.  In the latter interesting case, the unidirectional transport of light is achieved even in the absence of self-trapping nonlinearity, as shown below.
\par Below we present the dynamics of such interesting case,
\begin{figure}[]
\begin{center}
   \includegraphics[width=1.1\linewidth]{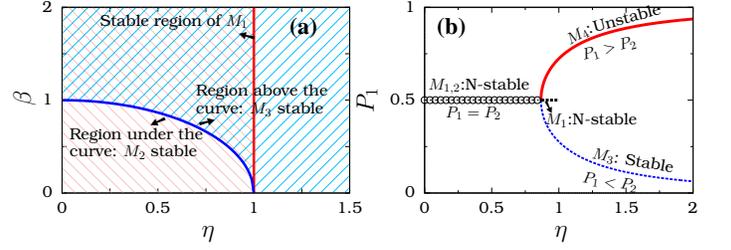}
\end{center}
   \caption{(color online) Fig.(a): Neutrally stable/stable regions of symmetric and asymmetric modes of system (\ref{non2}) for $k=0.5$, $P(0)=P=1.0$. Fig.(b) Bifurcation of asymmetric modes with opposite stabilities and is plotted for $\beta=0.5,$ $k=0.5$ and $P=1.0$.}
\label{f5}   
\end{figure}
  \begin{eqnarray}
i \dot{\phi_1}= - \beta |\phi_1|^2 \phi_1 - k \phi_2 +i \eta |\phi_2|^2 \phi_1, \nonumber \\
i \dot{\phi_2}= - \beta |\phi_2|^2 \phi_2 - k \phi_1 -i \eta |\phi_1|^2 \phi_2.
\label{non2}
\end{eqnarray}
\par  First let us study the dynamics and non-reciprocal nature of the system. For this purpose the equations corresponding to the amplitude and the phase difference of the system are
\begin{eqnarray}
\dot{R_1}&=&k R_2 \sin \delta+\eta R_1 R_2^2, \label{e1_3} \\
\dot{R_2}&=&-k R_1 \sin \delta- \eta R_1^2 R_2, \label{e2_3} \\
\dot{\delta}&=&(\beta R_1 R_2 - k \cos \delta)(R_1^2 - R_2^2).
\label{e3_3}
\end{eqnarray}
\par From the conservative nature of the system, we can reduce the order of the above equation through the substitution of $R_2= \sqrt{P-R_1^2}$. Thus the above equations will reduce to
\begin{eqnarray}
\dot{R_1}&=&(k \sin \delta+\eta R_1 \sqrt{P-R_1^2}) \sqrt{P-R_1^2}, \label{sec1_3} \\
\dot{\delta}&=&(\beta R_1 \sqrt{P-R_1^2} - k \cos \delta)\frac{(2 R_1^2-P)}{R_1 \sqrt{P-R_1^2}},
\label{sec2_3}
\end{eqnarray}
While tracing out the stationary modes of light corresponding to $\dot{R_1}=\dot{\delta}=0$, we can find the symmetric modes of light are given by 
\begin{eqnarray}
M_1: R_1^*=\frac{P}{2}, \; \cos \delta^*= - \frac{\sqrt{4 k^2- \eta^2 P^2}}{2 k},  \label{mm1_3} \\
M_2: R_1^*=\frac{P}{2},\; \cos \delta^*= + \frac{\sqrt{4 k^2- \eta^2 P^2}}{2 k},  \label{mm2_3} \\
\end{eqnarray} 
where in both cases $\sin \delta^*=-\frac{\eta P}{2 k}$ (Note that $\sin \delta^*=\frac{\eta P}{2 k}$, leads to a negative $R_1^*$ which is physically inadmissible and so this case is not considered).   These modes exist only when $4 k^2-\eta^2 P^2>0$.    Similarly the asymmetric modes take the form 
\begin{eqnarray}
M_3: R_1^*=\sqrt{\frac{P+\sqrt{P^2-\frac{4 k^2}{(\beta^2+\eta^2)}}}{2}}, \label{mm3_3} \\
M_4: R_1^*=\sqrt{\frac{P-\sqrt{P^2-\frac{4 k^2}{(\beta^2+\eta^2)}}}{2}}, \label{mm4_3} 
\end{eqnarray}
where $\cos \delta^*=\frac{\beta}{\sqrt{\beta^2+\eta^2}}$ and $\sin \delta^*=\frac{-\eta}{\sqrt{\beta^2+\eta^2}} $. One can find that these asymmetric modes exist only when $\beta^2+\eta^2> \frac{4 k^2}{P^2}$. 
\par By analyzing the stability of each one of the modes, one can find that the eigenvalues corresponding to the symmetric and asymmetric modes are 
\begin{small}
\begin{eqnarray}
M_1: \lambda&=&\pm i \sqrt{ \sqrt{4 k^2-P^2 \eta^2}(\sqrt{4 k^2- P^2 \eta^2}+\beta P)},\quad \label{lam1_3} \\
M_2: \lambda&=&\pm i \sqrt{\sqrt{4 k^2-P^2 \eta^2}(\sqrt{4 k^2- P^2 \eta^2}-\beta P)}, \quad \label{lam2_3} \\
M_3: \lambda&=&(- \eta \pm i \beta) \sqrt{P^2-\frac{4 k^2}{\beta^2+\eta^2}},  \quad \label{lam3_3} \\
M_4: \lambda&=& (\eta \pm i \beta) \sqrt{P^2-\frac{4 k^2}{\beta^2+\eta^2}}.
\quad \label{lam4_3} 
\end{eqnarray}
\end{small}
\par From the above, one can find that the symmetric mode $M_1$ is neutrally stable in the region $\eta< \frac{2 k}{P}$, whereas the other symmetric mode ($M_2$) is neutrally stable only when $\beta<\frac{\sqrt{4 k^2- P^2 \eta^2}}{P}$ besides $\eta<\frac{2k}{P}$.  Looking onto the eigenvalues of the asymmetric modes $M_3$ and $M_4$,  the mode $M_3$ is stable for $\beta^2+\eta^2> \frac{4 k^2}{P^2}$ whereas the mode $M_4$ is unstable for all values of parameters.  
\par The regions in which each of the modes are stable are given in Fig. \ref{f5}(a).   The figure clearly shows that the increase in either $\beta$ or $\eta$ leads to the stabilization of asymmetric fixed points which results in spontaneous symmetry breaking.  Also the bifurcation that stabilizes the asymmetric mode is like a tangent bifurcation where the bifurcation gives rise to two stationary modes ($M_3$ and $M_4$) with opposite stabilities. The bifurcation diagram of the case has also been given in Fig. \ref{f5}(b).  A closer look at such bifurcation of asymmetric fixed points reveals that it denotes the emergence of non-reciprocal nature.  This is because that unlike the earlier conservative cases discussed in Sec. II, here only one of the asymmetric modes $M_3$ with $R_1^* > R_2^*$ is stable (where the power in the first and second waveguides are related to $R_1^*$ and $R_2^*$ as $P_1={R_1^*}^2$ and $P_2={R_2^*}^2$). This implies that independent of whether the input power is given in the first or second waveguide, $M_3$ becomes stable and gives rise to localization of light in the first waveguide.  In the earlier conservative cases both the asymmetric modes are stable and depending on the input power distribution, the light is localized in one of the waveguides. Thus the above results show the unidirectional propagation of light or nonreciprocal nature of light in the present case.  One can also find that the dynamics of the system in both the unbroken and broken regions resemble that of the one shown in Fig. \ref{f12}.  Also it is clear from Fig. \ref{f5}(a) that the non-reciprocal nature arises even in the case $\beta=0$.

\section{\label{sum}Summary}
\par  In this article, we have studied the dynamics of a class of $\cal{PT}$ symmetric systems in which the coupling is responsible for the $\cal{PT}$ nature.  Importantly, we investigated the situations that can be used for the construction of unidirectional devices such as optical diodes.  We found that the $\cal{PT}$ symmetric systems showing spontaneous symmetry breaking through tangent bifurcation are found to be non-reciprocal and helpful in the construction of unidirectional transport devices.  The systems that show symmetry breaking through pitchfork bifurcation are found to be reciprocal, except for the cases with nonlinear interaction of the form $\phi_2^{n+1} {\phi_1^*}^n$ with $n=1,2$.  In the latter cases, although the systems show non-reciprocal phase shift, the symmetry is broken through pitchfork bifurcation.  Excluding this type of nonlinearity ($\phi_2^{n+1} {\phi_1^*}^n$), we summarize the conclusions below. 
\par{\it Linear $\cal{PT}$ coupling:} In the cases with linear $\cal{PT}$ coupling, all the systems having {\it $\cal{PT}$-term + self trapping nonlinearity + non-conservative nature} are found to be non-reciprocal.   Considering the conservative cases, the systems having nonlinearities of the form $G = |\phi_2|^2 \phi_1^2 \phi_2^*$ and $|\phi_1|^2 |\phi_2|^2 \phi_2$ are found to be non-reciprocal and all the other systems are reciprocal.  The non-reciprocal systems admit unidirectional transport of light when the $\cal{PT}$ symmetry of the system is spontaneously broken.  In this connection, except for the non-conservative nonlinearity case $|\phi_1|^2 \phi_1^2 \phi_2^*$, the spontaneous symmetry breaking can be seen in all the non-reciprocal systems and so they are found to be useful for the construction of unidirectional transport devices.  We also note that in the absence of self-trapping nonlinearity, many of these systems lose their non-reciprocal nature and in none of the cases the unidirectional transportation has been achieved.
\par {\it Nonlinear $\cal{PT}$ coupling:} In the case of nonlinear $\cal{PT}$ coupling systems, we find that all the conservative and non-conservative systems show non-reciprocal dynamics in the presence of self-trapping nonlinearities.  Except for the cases with nonlinearities  $G=|\phi_2|^4 \phi_1$ and  $|\phi_1|^2 \phi_1^2 \phi_2^*$, in all the other conservative and non-conservative systems the spontaneous $\cal{PT}$ symmetry breaking can be seen and so all these systems are useful for the construction of unidirectional transport devices in the presence of self-trapping nonlinearity.   We also note that even in the absence of self-trapping terms, for the odd case of nonlinear coupling $G=|\phi_2|^2\phi_1$ unidirectional transport of light is possible.\\ 
\par The above obtained results have been verified with nonlinearities upto order five.  Further, the efforts to engineer such models with gyro-magnetic or magneto-optic materials and the experimental observation on light localization in such models may reveal the adaptability and applicability of these models to real world systems.

\section*{Acknowledgement}
SK thanks the Department of Science and Technology (DST), Government of India, for providing a INSPIRE Fellowship.  The work of VKC is supported by the SERB-DST Fast Track scheme for young scientists under Grant No.YSS/2014/000175.  The work of MS forms part of a research project sponsored by Department of Science and Technology, Government of India.  The work of ML is supported by a NASI Senior Scientist Platinum Jubilee fellowship program.   
\appendix

\section{\label{ap1}Analytical results for $G=|\phi_1|^2 |\phi_2|^2 \phi_2$} 
We consider $G$ in (\ref{gcon}) to be $|\phi_1|^2 |\phi_2|^2 \phi_2$ so that the coupled mode equation turn out to be of the form
\begin{eqnarray}
i \frac{d \phi_1}{dz}  &=& - \beta |\phi_1|^2 \phi_1 - k \phi_2 + i a \phi_2+ \alpha |\phi_1|^2 |\phi_2|^2 \phi_2, \quad \nonumber \\
i \frac{d \phi_2}{dz}  &=& - \beta |\phi_2|^2 \phi_2 - k \phi_1 - i a \phi_1+ \alpha |\phi_1|^2 |\phi_2|^2 \phi_1, \quad
\end{eqnarray}
where $\alpha$ is the strength of the considered nonlinear coupling term.  Similar to the case (\ref{lin1}), we try to find the equations for the amplitudes ($R_1$ and $R_2$) and phase difference ($\delta$) of the nonlinear modes of the system by utilizing the integral of motion of the system ($P=|\phi_1|^2 + |\phi_2|^2$). By doing so, we obtain
\begin{small}
\begin{eqnarray}
\dot{R_1}&=\left((k-\alpha R_1^2(P-R_1^2)) \sin \delta+a \cos \delta\right) \sqrt{P-R_1^2}, \quad \nonumber \\
\dot{\delta}&=\left[\beta R_1 \sqrt{P-R_1^2}-\left((k-\alpha R_1^2(P-R_1^2)) \cos \delta-a \sin \delta\right)\right] \nonumber \\ &\hspace{4.5cm} \times \left(\frac{(2 R_1^2 -P)}{R_1 \sqrt{(P-R_1^2)}}\right).  \quad
\label{r1d_n1}
\end{eqnarray}
\end{small}
Note that $R_2^2=P-R_1^2$. From the above, the nonlinear modes of the system have been determined and the symmetric nonlinear modes ($R_1^*=R_2^*$) are found to be
\begin{eqnarray}
M_1:\; R_1^*&=&\sqrt{P/2}, \;\; \cos \delta^*=\frac{-(k-\frac{\alpha P^2}{4})}{\sqrt{a^2+(k-\frac{\alpha P^2}{4})^2}}\quad  \label{mm1_ap1}\\
M_2: \; R_1^*&=&\sqrt{P/2},  \;\; \cos \delta^*=\frac{(k-\frac{\alpha P^2}{4})}{\sqrt{a^2+(k-\frac{\alpha P^2}{4})^2}} \quad\label{mm2_ap1}
\end{eqnarray}
where  $\sin \delta^*=\frac{-a \cos \delta^*}{(k-\frac{\alpha P^2}{4})}$ and $R_2^*=\sqrt{P-{R_1^*}^2}=\sqrt{P/2}$.  One can also find the existence of asymmetric nonlinear stationary modes of the form
\begin{eqnarray}
M_{3}:\;R_1^*&=&\frac{1}{\sqrt{2}}\sqrt{P+\sqrt{P^2-4 C_1^2}},   \label{mm3_ap1} \\
M_{4}: \;R_1^*&=&\frac{1}{\sqrt{2}}\sqrt{P-\sqrt{P^2-4 C_1^2}}. \label{mm4_ap1} 
\end{eqnarray}
with $\cos \delta=-\frac{(k-\alpha C_1^2)}{\sqrt{a^2+(k-\alpha C_1^2)^2}}$, $\sin \delta=\frac{a}{\sqrt{a^2+(k-\alpha C_1^2)^2}}$ and 
\begin{eqnarray}
C_1=\sqrt{\frac{(\beta^2+2 k \alpha) + \sqrt{(\beta^2+2 k \alpha)^2-4 \alpha^2(a^2+k^2)}}{2 \alpha^2}}. \quad \label{c1_ap1}
\end{eqnarray}
  Eq. (\ref{r1d_n1}) also has some more asymmetric modes with
\begin{eqnarray}
M_{5}:\;R_1^*&=\frac{1}{\sqrt{2}}\sqrt{P+\sqrt{P^2-4 C_2^2}},  \label{mm5_ap1} \\
M_{6}: \;R_1^*&=\frac{1}{\sqrt{2}}\sqrt{P-\sqrt{P^2-4 C_2^2}}. \label{mm6_ap1}  
\end{eqnarray}
where $\cos \delta=-\frac{(k-\alpha C_2^2)}{\sqrt{a^2+(k-\alpha C_2^2)^2}}$, $\sin \delta=\frac{a}{\sqrt{a^2+(k-\alpha C_2^2)^2}}$ and 
\begin{eqnarray}
C_2=\sqrt{\frac{(\beta^2+2 k \alpha) -\sqrt{(\beta^2+2 k \alpha)^2-4 \alpha^2(a^2+k^2)}}{2 \alpha^2}}. \qquad
\label{c2}
\end{eqnarray}
From (\ref{mm3_ap1}-\ref{mm6_ap1}), it is obvious that in the modes $M_3$ and $M_5$, $P_1>P_2$ where $P_1 (=R_1^2)$ and $P_2(=R_2^2)$ are powers in the first and second waveguides.  In the modes $M_4$ and $M_6$, $P_1<P_2$. 
\par We have studied the stability of the obtained modes and we find that the stability determining eigenvalues corresponding to the symmetric modes are
\begin{small}
\begin{align}
&M_1: \lambda=\pm i \left(\sqrt{a^2+\left(k-\frac{\alpha P^2}{4}\right)^2}\sqrt{\frac{P}{2}}\right)^{\frac{1}{2}} \qquad \quad \nonumber \\
& \qquad \qquad \left(\sqrt{a^2+\left(k-\frac{\alpha P^2}{4}\right)^2}+ \frac{\beta P}{2}\right)^{\frac{1}{2}} \\
&M_2: \lambda=\pm i \left(\sqrt{a^2+\left(k-\frac{\alpha P^2}{4}\right)^2}\sqrt{\frac{P}{2}}\right)^{\frac{1}{2}} \qquad \quad \nonumber \\
&\qquad \qquad \left(\sqrt{a^2+\left(k-\frac{\alpha P^2}{4}\right)^2}- \frac{\beta P}{2}\right)^{\frac{1}{2}}
\label{eig_m1m3}
\end{align}
\end{small}
In the above, the mode $M_1$ is found to be neutrally stable for all positive values of the parameters and the mode $M_2$ is neutrally stable for the values of $\beta< {\frac{2}{P}}\sqrt{a^2+\left(k-\frac{\alpha P^2}{4}\right)^2}.$  The eigenvalues of the asymmetric modes $M_3$ and $M_4$ are
\begin{small}
\begin{align}
&M_3: \lambda=-\frac{2a \alpha C_1 \sqrt{P^2-4 C_1^2}}{\tilde{a_1}}\pm \left[\left(\frac{2a \alpha C_1 \sqrt{P^2-4 C_1^2}}{\tilde{a_1}} \right)^2 \right. \quad \quad \nonumber \\
& \quad \quad \left.  -4\left(\beta \tilde{a_1}+2 \alpha C_1 (k-\alpha C_1^2)\right) \frac{P^2-4 C_1^2}{C_1}\right]^{\frac{1}{2}} \label{eig_m3ap} \\
&M_4: \lambda=+\frac{2a \alpha C_1 \sqrt{P^2-4 C_1^2}}{\tilde{a_1}}\pm \left[\left(\frac{2a \alpha C_1 \sqrt{P^2-4 C_1^2}}{\tilde{a_1}} \right)^2 \right. \quad \quad \nonumber \\
& \quad \quad \left.  -4\left(\beta \tilde{a_1}+2 \alpha C_1 (k-\alpha C_1^2)\right) \frac{P^2-4 C_1^2}{C_1}\right]^{\frac{1}{2}}, 
\label{eig_m4ap}
\end{align}
\end{small}
where $\tilde{a_1}=\sqrt{a^2+(k-\alpha C_1^2)^2}$.
Note that whenever the term inside the square root in (\ref{eig_m3ap}) and (\ref{eig_m4ap}) is lesser than $\frac{2a \alpha C_1 \sqrt{P^2-4 C_1^2}}{\tilde{a_1}}$ or becomes negative, the mode $M_3$ becomes stable.  But the mode $M_4$ will never be stable (the parameters are considered to be positive). Thus in contrast to the case (\ref{lin1}), there occurs tangent like bifurcation which give rise two asymmetric modes $M_3$ and $M_4$ with opposite stabilities.  Here the stabilization of $M_3$ denotes $P_1>P_2$, thus there is a possibility to localize the light in first waveguide and to achieve unidirectional transport of light. 
\par Considering the other asymmetric modes $M_5$ and $M_6$, their eigenvalues are given by
\begin{small}
\begin{align}
&M_5:& \lambda=-\frac{2a \alpha C_2 \sqrt{P^2-4 C_2^2}}{\tilde{a_2}}\pm \left[\left(\frac{2a \alpha C_2 \sqrt{P^2-4 C_2^2}}{\tilde{a_2}} \right)^2 \right. \quad \quad \nonumber \\
&& \quad \quad \quad \left.  -4\left(\beta \tilde{a_2}+2 \alpha C_2 (k-\alpha C_2^2)\right) \frac{P^2-4 C_2^2}{C_2}\right]^{\frac{1}{2}} \\
&M_6:& \lambda=+\frac{2a \alpha C_2 \sqrt{P^2-4 C_2^2}}{\tilde{a_2}}\pm \left[\left(\frac{2a \alpha C_2 \sqrt{P^2-4 C_2^2}}{\tilde{a_2}} \right)^2 \right. \quad \quad \nonumber \\
&& \quad \quad \quad \left.  -4\left(\beta \tilde{a_2}+2 \alpha C_2 (k-\alpha C_2^2)\right) \frac{P^2-4 C_2^2}{C_2}\right]^{\frac{1}{2}}
\label{eig_m3m6}
\end{align}
\end{small}
where $\tilde{a_2}=\sqrt{a^2+(k-\alpha C_2^2)^2}$.
Similar to the previous case, here also one of the modes $M_5$ ($P_1>P_2$) becomes stable and the other mode $M_6$ ($P_1<P_2$) is unstable.
\par Also one can note that all the asymmetric modes will not exist for $\beta=0$.  Thus the symmetry will not be broken for $\beta=0$.


\begin{thebibliography}{99}
\bibitem{bendr}C. M. Bender and S. Boettcher, Phys. Rev. Lett, {\bf 80} 5243 (1998). 
\bibitem{r3} Z. H. Musslimani, K. G. Makris, R. El-Ganainy and D. N. Christodoulides, Phys. Rev. Lett. {\bf 100} 030402 (2008).
\bibitem{r4} K. G. Makris, R. El-Ganainy, D. N. Christodoulides and Z. H. Musslimani, Phys. Rev. Lett. {\bf 100} 103904 (2008). 
\bibitem{r41} A. Guo, G. J. Salamo, D. Duchesne, R. Morandotti, M. Volatier-Ravat, V. Aimez, G.A. Siviloglou, and D. N. Christodoulides, Phys Rev. Lett. {\bf 103} 093902 (2009).
\bibitem{r42} C. E. R\"uter, K. G. Makris, R. El-Ganainy, D. N. Christodoulides, M. Segev, and D. Kip, Nature Phys. {\bf 6} 192 (2010).
\bibitem{r43} A. Regensburger, C. Bersch, M. A. Miri, G. Onishchukov, D. N. Christodoulides, and U. Peschel, Nature {\bf 488} 167 (2012).
\bibitem{r5} H. Benisty \emph{et al}. Opt. Express {\bf 19} 18004 (2011).
\bibitem{r55} A. Lupu, H. Bensity and A. Degiron, Opt. Express {\bf 21} 21651 (2013).
\bibitem{r6} C. Hang, G. Huang and V. V. Konotop, Phys. Rev. Lett.  {\bf 110} 083604 (2013).
\bibitem{r7} J. Sheng, M. A. Miri, D. N. Christodoulides and M. Xiao, Phys. Rev. A {\bf 88} 041803 (2013).
\bibitem{r8} M. Kreibich, J. Main, H. Cartarius, and G. Wunner, Phys. Rev. A {\bf 93} 023624 (2016).
\bibitem{r9} H. Cartarius and G. Wunner, Phys. Rev. A {\bf 86} 013612 (2012).
\bibitem{acous} X. Zhu, H. Ramezani, C. Shi, J. Zhu and X. Zhang, Phys. Rev. X {\bf 4} 031042 (2014).
\bibitem{r11} J. Schindler, A. Li, M. C. Zheng, F. M. Ellis and T. Kottos, Phys. Rev. A {\bf 84} 040101 (2011).
\bibitem{r12}N. Bender, S. Factor,  J. D. Bodyfelt,  H. Ramezani, D. N. Christodoulides, F. M. Ellis,  and T. Kottos, Phys. Rev. Lett. {\bf 110} 234101 (2013).
\bibitem{r13}C. M. Bender, M. Gianfreda, S. K. \"{O}zdemir, B. Peng, and L. Yang, {Phys. Rev. A} {\bf 88} 062111 (2013).
\bibitem{r14}J. Cuevas, P.G. Kevrekidis, A. Saxena and A. Khare, {Phys. Rev. A} {\bf 88} 032108 (2013).
\bibitem{r15}S. Karthiga, V. K. Chandrasekar, M. Senthilvelan and M. Lakshmanan, Phys. Rev. A {\bf 93} 012102  (2016).
\bibitem{nature} L. Chang et al, Nature Photonics, {\bf 8} 524  (2014).
\bibitem{ref2} H. Ramezani, T. Kottos, R. El-Ganainy, and D. N. Christodoulides, Phys. Rev. A {\bf 82} 043803 (2010). 
\bibitem{bec} H. Cartarius and G. Wunner, Phys. Rev. A {\bf 86}  013612 (2012), 
\bibitem{kevre}K. Li and P. G. Kevrekidis, Phys. Rev. E {\bf 83} 066608 (2011).
\bibitem{kivsh}A. A. Sukhorukov, Z. Xu, and Y. S. Kivshar, Phys. Rev. A {\bf 82} 043818 (2010).
\bibitem{flach}I. V. Barashenkov, G. S. Jackson and S. Flach, Phys. Rev. A {\bf 88} 053817 (2013).
\bibitem{susan}J. Pickton and H. Susanto, Phys. Rev. A {\bf 88} 063840 (2013).
\bibitem{bar1}I. V. Barashenkov, Phys. Rev. A. {\bf 90} 045802 (2014). 
\bibitem{bar2} I. V. Barashenkov, D. E. Pelinovsky and P. Dubard, J. Phys. A: Math. Theor. {\bf 48} 325201 (2015). \bibitem{bar3}I. V. Barashenkov, M. Gianfreda, J. Phys. A: Math. Theor. {\bf 47} 282001 (2014).
 \bibitem{genr}D. E. Pelinovsky, D. A. Zezyulin and V. V. Konotop, J. Phys. A: Math. Theor. {\bf 47}  085204 (2014).
\bibitem{bors}A. E. Miroshnichenko, B. A. Malomed, and Y. S. Kivshar, Phys. Rev. A {\bf 84} 012123 (2011).
\bibitem{circ}W. Qiu, Z. Wang, and M. Soljaci\'c, Opt. Express {\bf 19} 22248 (2011).
\bibitem{yar_p} A. Yariv, IEEE J. Quant. Electron. {\bf 9} 919 1973.
\bibitem{yariv} A. Yariv, \emph{Quantum Electronics}, (John Wiley \& sons, USA, 1987).
\bibitem{compact}A. V. Yulin and V. V. Konotop, Opt. Lett. {\bf 88} 4880 (2013).
\bibitem{kev}H. Xu, P. G. Kevrekidis and A. Saxena, J. Phys. A. Math. Theor. {\bf 48} 055101 (2015).
\bibitem{hamel}P. Hamel, S. Haddadi, F. Raineri, P. Monnier, G. Beaudoin, I. Sagnes, A. Levenson, and A. M. Yacomotti, Nature Photonics {\bf 9} 311 (2015).
\bibitem{malo} B. A. Malomed, Nature Photonics {\bf 9} 287 (2015).
 
\end{thebibliography}
\end{document}